# On Linear Index Coding for Random Graphs

Ishay Haviv[*]      Michael Langberg[†]

**Abstract**

A sender wishes to broadcast an $n$ character word $x \in \mathbb{F}^n$ (for a field $\mathbb{F}$) to $n$ receivers $R_1, \ldots, R_n$. Every receiver has some side information on $x$ consisting of a subset of the characters of $x$. The side information of the receivers is represented by a graph $G$ on $n$ vertices in which $\{i, j\}$ is an edge if and only if $R_i$ knows $x_j$. In the *index coding* problem the goal is to encode $x$ using a minimum number of characters in $\mathbb{F}$ in a way that enables every $R_i$ to retrieve the $i$th character $x_i$ using the encoded message and the side information. An index code is *linear* if the encoding is linear, and in this case the minimum possible length is known to be equal to a graph parameter called *minrank* (Bar-Yossef et al., FOCS'06). Several bounds on the minimum length of an index code for side information graphs $G$ were shown in the study of index coding by Bar-Yossef et al. (FOCS'06), Lubetzky and Stav (FOCS'07), Alon et al. (FOCS'08), and Blasiak et al. (manuscript, arXiv'10). However, the minimum length of an index code for the random graph $G(n, p)$ is far from being understood.

In this paper we initiate the study of the *typical* minimum length of a *linear* index code for $G(n, p)$ over a field $\mathbb{F}$. First, we prove that for every constant size field $\mathbb{F}$ and a constant $p$, the minimum length of a linear index code for $G(n, p)$ over $\mathbb{F}$, i.e., the minrank of $G(n, p)$ over $\mathbb{F}$, is *almost surely* $\Omega(\sqrt{n})$. Second, we introduce and study the following two restricted models of index coding:

1. A *locally decodable index code* is an index code in which the receivers are allowed to query at most $q$ characters from the encoded message. We prove that the minimum length of a linear locally decodable index code for $G(n, p)$ over $\mathbb{F}$ with $q$ queries is almost surely $\widetilde{\Omega}(\frac{n}{q})$ assuming that $q = \widetilde{\Omega}(n^{\frac{1}{3}})$. In particular, for locally decodable index codes with some $q = o(\sqrt{n})$ we get an $\omega(\sqrt{n})$ lower bound.

2. A *low density index code* is a linear index code in which every character of the word $x$ affects at most $q$ characters in the encoded message. Equivalently, it is a linear code whose generator matrix has at most $q$ nonzero entries in each row. We prove that in order to show an $\omega(\sqrt{n})$ lower bound on the minimum length of a linear index code for $G(n, p)$ over $\mathbb{F}$ it suffices to show such a lower bound on the length of a *low density* index code for $G(n, p)$ over $\mathbb{F}$ with *some* $q = \omega(1)$. In addition, we prove that the minimum length of a low density index code for $G(n, p)$ over $\mathbb{F}$ is almost surely at least $n^{1-\varepsilon}$ for $q = 2$ and at least $n^{\frac{2}{3}-\varepsilon}$ for $q = 3$ for any sufficiently small constants $p$ and $\varepsilon > 0$.

---

[*]The Blavatnik School of Computer Science, Tel Aviv University, Tel Aviv 69978, Israel. Supported by the Adams Fellowship Program of the Israel Academy of Sciences and Humanities. Work done in part while at the Open University of Israel.

[†]The Computer Science Division, Open University of Israel, Raanana 43107, Israel. Email: mikel@openu.ac.il. Work supported in part by ISF grant 480/08 and by the Open University of Israel's research fund (grant no. 46114).

# 1 Introduction

In the *index coding* problem, a sender wishes to broadcast an $n$ character word $x \in \mathbb{F}^n$ (for a finite field $\mathbb{F}$) to $n$ receivers $R_1, \ldots, R_n$ in a way that enables every $R_i$ to retrieve the $i$th character $x_i$. Every receiver has some side information on $x$. The side information is represented by a directed graph $G$ on the vertex set $[n] = \{1, 2, \ldots, n\}$ in which a vertex $i$ is connected to a vertex $j$ if and only if the receiver $R_i$ knows $x_j$. Given a side information graph $G$, the goal is to find a coding scheme of minimum length, by which every receiver $R_i$ is able to retrieve $x_i$ given the encoded message and the side information that it has on $x$ according to $G$. The settings are naturally extended to undirected graphs in which an edge $\{i, j\}$ meas that $R_i$ knows $x_j$ and $R_j$ knows $x_i$.

For example, assume that every receiver $R_i$ knows $x_j$ for every $j \in [n] \setminus \{i\}$. The corresponding side information graph is the complete graph on the vertex set $[n]$. In this case, broadcasting the sum $\sum_{i \in [n]} x_i$ over $\mathbb{F}$ enables every receiver $R_i$ to retrieve $x_i$, and hence the minimum message length required here is 1.

The study of index coding was initiated by Birk and Kol in [6] and further developed by Bar-Yossef, Birk, Jayram and Kol in [5]. This research is motivated by applications, such as video on demand and wireless networking, in which a network transmits information to clients, and during the transmission every client misses some of the information. At this step, the clients have side information on the transmitted information, and the network is interested in minimizing the broadcast length in a way that enables the clients to decode their target (see, e.g., [23]).

Research on index coding is motivated by several questions in theoretical computer science. For example, index coding is a natural version of the *one-way communication complexity* problem of the *indexing* function studied in [14]. In this problem, Alice is given an $n$ bit string $x$, sends a single message to Bob, and Bob, given an index $i$, should be able (possibly probabilistically) to discover $x_i$. The goal is to minimize the length of Alice's message. The index coding problem over $\mathbb{F}_2$ is equivalent to this question once we restrict Bob to act deterministically and allow him to use some side information on $x$, depending on $i$. The study of index coding is also motivated by the more general problem of *network coding*, introduced by Ahlswede et al. [1]. El Rouayheb et al. showed in [10] that network coding instances can be efficiently reduced to index coding instances. Hence, understanding index coding capacities is motivated by applications in computational complexity regarding deciding and approximating the network coding problem (see [16, 15]).

For a graph $G$ and a field $\mathbb{F}$ we denote by $\beta_1(G)$ the minimum length of an index code for $G$ over $\mathbb{F}$. This graph parameter is well-known to be related to several classical graph parameters. Indeed, for an undirected graph $G$, $\beta_1(G)$ is bounded from below by $\alpha(G)$, the maximum size of an independent set in $G$, as follows from the fact that an independent set in $G$ corresponds to a set of receivers with no mutual information. On the other hand, $\beta_1(G)$ is bounded from above by $\chi(\overline{G})$, the clique cover number of $G$, as follows from broadcasting the sum over $\mathbb{F}$ of the characters corresponding to the vertices in every clique in an optimal clique cover.

In [5], Bar-Yossef et al. identified an algebraic graph parameter, called *minrank*[1] and denoted by $\mathrm{minrk}_{\mathbb{F}}(G)$, which upper bounds $\beta_1(G)$. Interestingly, they proved that the minrank of a graph $G$ over $\mathbb{F}$ equals the minimum length of a *linear* index code for $G$ over $\mathbb{F}$ (i.e., an index code whose encoding function is linear). They also proved that their upper bound is tight and is equal to $\beta_1(G)$ for several graph families for the binary field $\mathbb{F}_2$. This includes directed acyclic graphs, perfect graphs, odd holes (undirected odd-length cycles of length at least 5) and odd anti-holes (complements of odd holes). These results raised the question whether the minrank parameter characterizes the minimum length of general index codes. This question was answered in the negative by Lubetzky and Stav [18], who showed that for any $\varepsilon > 0$ and a sufficiently large $n$ there is an $n$ vertex graph $G$ with $\beta_1(G) \leq n^\varepsilon$ and $\text{minrk}_{\mathbb{F}_2}(G) \geq n^{1-\varepsilon}$ (see [3] for additional counterexamples). We note that the proof in [18] uses a property of the minrank (see also [12]), saying that for every field $\mathbb{F}$ and an $n$ vertex undirected graph $G$,

$$\text{minrk}_{\mathbb{F}}(G) \cdot \text{minrk}_{\mathbb{F}}(\overline{G}) \geq n. \quad (1)$$

---
[1]See Definition 2.1.

The first to define the minrank parameter was Haemers [11, 12], who related it to what is known as the Shannon capacity of graphs introduced in [22]. Haemers showed that for every field $\mathbb{F}$ and an undirected graph $G$, $\alpha(G) \leq c(G) \leq \text{minrk}_{\mathbb{F}}(G)$, where $c(G)$ stands for the Shannon capacity of $G$. He also showed that there are graphs for which the minrank upper bound on the Shannon capacity is tighter than the one given by the well-known Lovász $\theta$-function introduced in [17]. We note that calculating the minrank of a given input graph is known to be NP-hard [21], as opposed to the efficiently computable Lovász $\theta$-function.

The following theorem summarizes some of the bounds mentioned above.

**Theorem 1.1** ([11, 12, 5]). *For every field $\mathbb{F}$ and an undirected graph $G$,*

$$\alpha(G) \leq \beta_1(G) \leq \text{minrk}_{\mathbb{F}}(G) \leq \chi(\overline{G}).$$

All the inequalities in the above statement are known to be strict for certain graphs. This makes the task of understanding $\beta_1(G)$ challenging. A fundamental parameter to study in this context is the typical value of $\beta_1(G)$ for *random graphs* $G$. This question was raised by Lubetzky and Stav in [18] for the well-known random graph $G(n, \frac{1}{2})$, where $G(n, p)$ denotes the random undirected graph with $n$ vertices and edge probability $p$. In this paper we focus on linear index codes and study the following question:

*What is the typical minimum length of a* linear *index code for the random graph $G(n, p)$ over $\mathbb{F}$?*

Equivalently, we are asking for the typical minrank over $\mathbb{F}$ of the random graph $G(n, p)$.

Let us start with some bounds yielded by Theorem 1.1. Both the independence number and the clique cover number of $G(n, p)$ are well understood (see [9] for the former and [8, 19] for the latter). For a constant edge probability $p$, we obtain that almost surely (i.e., with probability that tends to 1 as $n$ tends to infinity),

$$(1 \pm o(1)) \cdot \frac{2 \log n}{\log \frac{1}{1-p}} \leq \text{minrk}_{\mathbb{F}}(G(n,p)) \leq (1 \pm o(1)) \cdot \frac{n \log \frac{1}{p}}{2 \log ((1-p)n)}.$$

In short, for a constant $p$, almost surely, $\Omega(\log n) \leq \text{minrk}_{\mathbb{F}}(G(n,p)) \leq O(\frac{n}{\log n})$. The gap between these lower and upper bounds is exponential, and, surprisingly, no better bounds are known to hold almost surely for $G(n, p)$. Yet, it is plausible to expect the minrank of $G(n, p)$ to be much higher than the $\Omega(\log n)$ lower bound, since the bound in (1) implies that the *expected* minrank of $G(n, p)$ is $\Omega(\sqrt{n})$ for $p = \frac{1}{2}$ (and hence for any $p \leq \frac{1}{2}$ as well). To see this, notice that if $G$ is distributed according to $G(n, \frac{1}{2})$ then so is its complement, and hence the probability that $\text{minrk}_{\mathbb{F}}(G) \geq \sqrt{n}$ is at least $\frac{1}{2}$. We note, though, that any $\omega(\sqrt{n})$ lower bound on the expectation above would imply an $\omega(\sqrt{n})$ lower bound which holds almost surely, as follows from the large deviation inequality for vertex exposure martingale (see, e.g., [4], Chapter 7). Understanding the true value of $\text{minrk}_{\mathbb{F}}(G(n,p))$ and, more specifically, the question whether one can show an $\omega(\sqrt{n})$ lower bound on it, are the driving force of this work.

## 1.1 Our Contribution

In the current paper we study the typical minimum length of a linear index code for the random graph $G(n,p)$ over a field $\mathbb{F}$. We start by showing that an $\Omega(\sqrt{n})$ lower bound holds with probability that (exponentially) tends to 1 as $n$ tends to infinity (and not only in expectation). In addition, the bound holds for every constant size field $\mathbb{F}$ and a constant edge probability $p$.[2]

**Theorem 1.2.** *For every constant size field $\mathbb{F}$ and a constant $p \in (0,1)$, almost surely*

$$\mathrm{minrk}_{\mathbb{F}}(G(n,p)) = \Omega(\sqrt{n}).$$

Observe that Theorem 1.2 implies that the random graph $G(n, \frac{1}{2})$ almost surely has an exponential gap between its independence number and its minrank over any constant size field. In [2], Alon conjectured that the Shannon capacity of $G(n, \frac{1}{2})$ satisfies $c(G(n, \frac{1}{2})) = O(\log n)$ almost surely. This, if true, would imply an exponential gap between the Shannon capacity and the minrank upper bound of Haemers [12] on it for a typical graph $G(n, \frac{1}{2})$.

In the attempt to understand where the minrank of $G(n,p)$ exactly lies in the range from $\sqrt{n}$ to $\frac{n}{\log n}$ we introduce and study two natural restricted models of index coding.

**Locally decodable index coding.** In our first model we study index codes in which the decoders are allowed to query a limited number of characters from the encoded message. More precisely, these are index codes in which the sender maps $x \in \mathbb{F}^n$ to an encoded message, and each of the receivers should be able to recover $x_i$ using at most $q$ queries to the encoded message and the information that the receiver has on $x$ according to the side information graph. The following theorem says that every linear locally decodable index code for $G(n,p)$ over $\mathbb{F}$ with $q$ significantly smaller than $\sqrt{n}$ almost surely has length much higher than $\sqrt{n}$. The $\widetilde{\Omega}$ notation is used to hide factors which are logarithmic in $n$.

**Theorem 1.3.** *For every constant size field $\mathbb{F}$ and a constant $p \in (0,1)$, if there exists a linear index code of length $\ell$ for $G(n,p)$ over $\mathbb{F}$, such that every decoding function queries at most $q = \widetilde{\Omega}(n^{\frac{1}{3}})$ characters from the encoded message, then almost surely $\ell = \widetilde{\Omega}(\frac{n}{q})$.*

We note that a locally decodable index code is a natural analogue of the widely studied object known as *locally decodable codes* introduced by Katz and Trevisan [13]. Roughly speaking, locally decodable codes enable a probabilistic decoding of any character of the original message by looking at a limited number of characters in a possibly corrupted encoded message.

**Low density index coding.** The second model we study consists of linear index codes in which every character of the word $x$ (that the sender wishes to broadcast) affects a limited number, say $q$, of characters in the encoded message. Such codes are generated by generator matrices in which every row has at most $q$ nonzero entries, thus we call them *low density generator matrix index codes* (or, in short, low density index codes).

Low density codes are usually not so useful in coding theory. The reason is that such codes have minimum distance at most $q$, whereas, in most applications, one desires codes of large minimum distance. However, for our purposes such codes turn to play a major role. More specifically, our next theorem says that improving the $\sqrt{n}$ lower bound on the length of low density index codes for $G(n,p)$ will imply such an improvement on the length of linear index codes for $G(n,p)$ in general. This is quite surprising since low density index codes intuitively seem significantly weaker than general linear index codes. We state this result here informally, and the formal statement can be found in Section 6.

---
[2]In fact, our proof provides a lower bound also for the case that $|\mathbb{F}|$ and $p$ depend on $n$. For the full statement of this theorem see Theorem 4.3.

**Theorem 1.4** (informal). *Assume that every linear index code for $G(n,p)$ over $\mathbb{F}$, with at most $q = \omega(1)$ nonzero entries in a row of its generator matrix, has length $\omega(\sqrt{n})$ with high probability. Then, almost surely,* $\mathrm{minrk}_\mathbb{F}(G(n,p)) = \omega(\sqrt{n})$.

Theorem 1.4 motivates studying lower bounds on the length of low density index codes for $G(n,p)$. Observe that the minimum length of a low density index code with $q = 1$ for a graph $G$ equals the clique cover number $\chi(\overline{G})$. This implies a tight lower bound of $\Omega(\frac{n}{\log n})$ for $q = 1$. We are also able to prove $\omega(\sqrt{n})$ lower bounds for low density index codes for $q = 2$ and $q = 3$, as stated below.

**Theorem 1.5.** *For every constant size field $\mathbb{F}$ and sufficiently small constants $\varepsilon, p > 0$,*

1. *every linear index code for $G(n,p)$ over $\mathbb{F}$, in which every character of the sent word affects at most 2 characters of the encoded message, almost surely has length at least $n^{1-\varepsilon}$, and*

2. *every linear index code for $G(n,p)$ over $\mathbb{F}$, in which every character of the sent word affects at most 3 characters of the encoded message, almost surely has length at least $n^{\frac{2}{3}-\varepsilon}$.*

## 1.2 Outline

The remainder of the paper is organized as follows. In Section 2 we provide some background preliminaries needed throughout the paper. In Section 3 we show that the minimum length of an index code for the (undirected) graph $G(n,p)$ is similar to that of directed random graphs. This enables us to simplify the presentation of our proofs by considering the directed random graph model. In Section 4 we prove the $\Omega(\sqrt{n})$ lower bound given in Theorem 1.2. In Section 5 we prove our result on locally decodable index codes, and in Section 6 we prove our results on low density index codes. The final Section 7 discusses some concluding remarks and open questions.

## 2 Preliminaries

In the index coding problem a sender wishes to broadcast a word $x \in \mathbb{F}^n$ (for a field $\mathbb{F}$) to $n$ receivers $R_1, \ldots, R_n$. Every receiver $R_i$ knows some fixed subset of the characters of $x$ and is interested solely in the character $x_i$. An $\ell$-*index code* for this setting is a length $\ell$ code over $\mathbb{F}$, which enables $R_i$ to recover $x_i$ for every $x \in \mathbb{F}^n$ and $i \in [n]$.

The index coding problem can be stated as a graph parameter. For a directed graph $G$ and a vertex $v$ let $N_G^+(v)$ denote the set of out-neighbors of $v$ in $G$, and for $x \in \mathbb{F}^n$ and $S \subseteq [n]$ let $x|_S$ denote the restriction of $x$ to the coordinates of $S$. The setting of the definition of an index code is characterized by the directed *side information graph* $G$ on the vertex set $[n]$ where $(i,j)$ is an edge if and only if the receiver $R_i$ knows $x_j$. An $\ell$-index code for $G$ over $\mathbb{F}$ is a function $E : \mathbb{F}^n \to \mathbb{F}^\ell$ and functions $D_1, \ldots, D_n$, so that for all $i \in [n]$ and $x \in \mathbb{F}^n$, $D_i(E(x), x|_{N_G^+(i)}) = x_i$. The definition of an index code is naturally extended to undirected graphs by replacing every undirected edge by two oppositely directed edges.

We say that the index code is *linear* if the encoding function $E$ is linear. It is not difficult to see that in the linear case it can be assumed, without loss of generality, that the linear function $E$ is homogenous. This means that there exist vectors $e_1, \ldots, e_\ell \in \mathbb{F}^n$ such that every $x \in \mathbb{F}^n$ is mapped by $E$ to $E(x) = (\langle e_1, x\rangle, \langle e_2, x\rangle, \ldots, \langle e_\ell, x\rangle)$. For example, for the binary field $\mathbb{F}_2$, every coordinate of $E(x)$ is the xor of a certain subset of the coordinates of $x$.

Bar-Yossef et al. [5] showed that the minimum length of a linear index code for $G$ over $\mathbb{F}$ equals $\text{minrk}_\mathbb{F}(G)$, a graph parameter defined as follows.

**Definition 2.1.** *Let $A = (a_{ij})$ be an n by n matrix over some field $\mathbb{F}$. We say that $A$ represents an $n$ vertex graph $G$ over $\mathbb{F}$ if $a_{ii} \neq 0$ for all $i$, and $a_{ij} = 0$ whenever $i \neq j$ and $(i,j)$ is not an edge in $G$. The minrank of a graph $G$ over $\mathbb{F}$ is defined as*

$$\text{minrk}_\mathbb{F}(G) = \min\{\text{rank}_\mathbb{F}(A) \mid A \text{ represents } G \text{ over } \mathbb{F}\}.$$

Let $E : \mathbb{F}^n \to \mathbb{F}^\ell$ be a linear $\ell$-index code for a graph $G$ and identify it with its generator matrix in $\mathbb{F}^{n \times \ell}$. Denote the $i$th column of $E$ by $e_i$ and denote $\text{span}(E) = \text{span}(e_1, \ldots, e_\ell)$. For a message $x \in \mathbb{F}^n$, the $i$th receiver is interested in $x_i$. In order to discover $x_i$ the $i$th receiver is allowed to use the codeword $E(x) = (\langle e_1, x\rangle, \langle e_2, x\rangle, \ldots, \langle e_\ell, x\rangle)$ and the side information that it has on $x$ according to $G$. It can be seen that the $i$th receiver is able to discover $x_i$ if and only if there exists a vector in $\text{span}(E)$ that is nonzero in the $i$th entry and is zero in all the entries that correspond to non-neighbors of $i$. This motivates the following definition which will be useful throughout the paper.

**Definition 2.2.** *For a graph $G$ on the vertex set $[n]$, a vector $v \in \mathbb{F}^n$ satisfies a vertex $i \in [n]$ if $v_i \neq 0$ and $v_j = 0$ for every $j \in [n] \setminus \{i\}$ such that $i$ is not connected to $j$ in $G$.*

Using this terminology, $E$ is a linear index code for $G$ if and only if every vertex $i \in [n]$ is satisfied by a vector in $\text{span}(E)$.

We need the following simple claim, in which we use $B_n(r)$ to denote the set of vectors in $\mathbb{F}^n$ of Hamming weight (i.e., number of nonzero entries) at most $r$.

**Claim 2.3.** *For every field $\mathbb{F}$, $n, \ell, r \in \mathbb{N}$, and a basis $E \in \mathbb{F}^{n \times \ell}$, the number of indices of coordinates that are nonzero in at least one vector in $\text{span}(E) \cap B_n(r)$ is at most $r \cdot \ell$.*

**Proof:** Consider the following process: start with $i = 1$, and at the $i$th step choose a vector $v_i \in \text{span}(E) \cap B_n(r)$ that has a nonzero coordinate which is zero in all the previously chosen vectors $v_1, \ldots, v_{i-1}$. Clearly, the process does not terminate as long as there is a coordinate that is nonzero in at least one vector in $\text{span}(E) \cap B_n(r)$ but is zero in all the chosen $v_i$'s. Observe that for every $i$, the vectors $v_1, \ldots, v_i$ are linearly independent. Therefore, the process terminates after at most $\ell$ steps. At each step we have at most $r$ new indices of nonzero coordinates since the $v_i$'s are in $B_n(r)$. This implies that the number of indices of coordinates that are nonzero in at least one vector in $\text{span}(E) \cap B_n(r)$ is at most $r \cdot \ell$. ∎

Let $G(n,p)$ denote the random *undirected* graph with $n$ vertices and edge probability $p$, and let $\vec{G}(n,p)$ denote the random *directed* graph with $n$ vertices and edge probability $p$. We say that $G(n,p)$, resp. $\vec{G}(n,p)$, satisfies a graph property *almost surely* if the probability that $G(n,p)$, resp. $\vec{G}(n,p)$, satisfies this property tends to 1 as $n$ tends to infinity.

Throughout the paper we ignore floors and ceilings whenever appropriate as this does not affect the asymptotic nature of our results.

## 3  $G(n,p)$ versus $\vec{G}(n,p)$

In this section we prove a lemma saying that the minimum length of an index code for $G(n,p)$ and the minimum length of an index code for $\vec{G}(n,p)$ behave similarly for constant edge probabilities.

We start with the intuitive proof idea and then turn to the proof. In the directed graph $\vec{G}(n,p)$ the probability that two vertices are connected by two oppositely directed edges is $p^2$. Hence, $\vec{G}(n,p)$ essentially contains a copy of $G(n,p^2)$. On the other hand, the probability that two vertices in $\vec{G}(n,p)$ are not connected at all is $(1-p)^2$, so $\vec{G}(n,p)$ is essentially contained in $G(n, 1-(1-p)^2)$. Therefore, a lower bound on the minimum length of an index code for $\vec{G}(n,p)$ for some constant $p$ implies a lower bound on that of $G(n,p')$ for some constant $p'$ and vice versa.

**Lemma 3.1.** *For every field $\mathbb{F}$, $n, \ell \in \mathbb{N}$ and $p \in (0,1)$, let*

$$P_1 = \Pr\left[\text{There exists an } \ell\text{-index code for } G(n, p^2) \text{ over } \mathbb{F}\right],$$
$$P_2 = \Pr\left[\text{There exists an } \ell\text{-index code for } \vec{G}(n, p) \text{ over } \mathbb{F}\right],$$
$$P_3 = \Pr\left[\text{There exists an } \ell\text{-index code for } G(n, p(2-p)) \text{ over } \mathbb{F}\right].$$

*Then, $P_1 \leq P_2 \leq P_3$.*
*In addition, the inequalities hold when we require the index codes in the three events to be linear, $(q, \ell)$-locally decodable for some $q \in \mathbb{N}$, or $(q, \ell)$-low density for some $q \in \mathbb{N}$.[3]*

**Proof:** We first show that $P_1 \leq P_2$. For a directed graph $G$, let $\widetilde{G}$ denote the undirected graph on the vertex set of $G$ in which two vertices are adjacent if and only if they are connected in $G$ by two oppositely directed edges. Observe that if $G$ is distributed according to $\vec{G}(n,p)$ then $\widetilde{G}$ is distributed according to $G(n, p^2)$. For a graph $G$ (directed or not) and $\ell \in \mathbb{N}$ we denote by $I_{G,\ell}$ the indicator variable of the event "There exists an $\ell$-index code for $G$ over $\mathbb{F}$". Notice that for every directed graph $G$ we have $I_{G,\ell} \geq I_{\widetilde{G},\ell}$, since every $\ell$-index code for $\widetilde{G}$ is an $\ell$-index code for $G$ as well. We get that

$$P_2 = \sum_{G \in \vec{G}(n,p)} I_{G,\ell} \cdot \Pr[G] \geq \sum_{G \in \vec{G}(n,p)} I_{\widetilde{G},\ell} \cdot \Pr[G] = \sum_{G' \in G(n,p^2)} I_{G',\ell} \cdot \Pr[G'] = P_1,$$

where the second equality holds since the probability of a graph $G'$ according to the distribution $G(n, p^2)$ equals the sum of the probabilities of all the graphs $G$ satisfying $\widetilde{G} = G'$ according to the distribution $\vec{G}(n,p)$.

The proof of the inequality $P_2 \leq P_3$ is similar. For a directed graph $G$, let $\widehat{G}$ denote the undirected graph on the vertex set of $G$ in which two vertices are adjacent if and only if they are connected in $G$ by at least one directed edge. Observe that if $G$ is distributed according to $\vec{G}(n,p)$ then $\widehat{G}$ is distributed according to $G(n, p(2-p))$ and that for every directed graph $G$ we have $I_{G,\ell} \leq I_{\widehat{G},\ell}$. We get that

$$P_2 = \sum_{G \in \vec{G}(n,p)} I_{G,\ell} \cdot \Pr[G] \leq \sum_{G \in \vec{G}(n,p)} I_{\widehat{G},\ell} \cdot \Pr[G] = \sum_{G' \in G(n,p(2-p))} I_{G',\ell} \cdot \Pr[G'] = P_3,$$

where the second equality holds since the probability of a graph $G'$ according to the distribution $G(n, p(2-p))$ equals the sum of the probabilities of all the graphs $G$ satisfying $\widehat{G} = G'$ according to the distribution $\vec{G}(n,p)$.

Finally, assume that the codes in the three events satisfy one (or more) of the properties mentioned in the statement of the lemma. It can be seen that an almost identical proof yields the result, since the inequalities $I_{\widetilde{G},\ell} \leq I_{G,\ell} \leq I_{\widehat{G},\ell}$ remain true when we require the code to have such a property in the definition of the event $I_{G,\ell}$. ∎

---

[3]See Definitions 5.1 and 6.1.

# 4 The $\Omega(\sqrt{n})$ Lower Bound

In this section we prove that $\text{minrk}_{\mathbb{F}}(G(n,p)) \geq \Omega(\sqrt{n})$ almost surely. By Lemma 3.1 it suffices to prove the lower bound for the directed random graph $\vec{G}(n,p)$. We start with some intuition. Fix a linear $\ell$-index code generated by $E \in \mathbb{F}^{n \times \ell}$ for certain $\ell = O(\sqrt{n})$. Our goal is to show that the probability that $E$ is an index code for $\vec{G}(n,p)$ is exponentially small, so that applying the union bound over all the codes $E$ will give us the result. As mentioned before, if $E$ is an index code for a graph on the vertex set $[n]$ then every vertex $i$ is satisfied by a vector in $\text{span}(E)$, i.e., there exists a vector $v \in \text{span}(E)$ such that $v_i \neq 0$ and $v_j = 0$ for all $j \neq i$ for which $(i,j)$ is not an edge in $G$. It is not hard to verify that any vector in $\text{span}(E)$ of Hamming weight $r$, whose $i$th entry is nonzero, satisfies a vertex $i$ with probability $p^{r-1}$. Using this, we show that the probability that at least $\frac{n}{2}$ vertices are satisfied by vectors of high Hamming weight is small (Lemma 4.1). On the other hand, we use Claim 2.3 to show that at most $\frac{n}{2}$ vertices can be satisfied by vectors of low Hamming weight (Lemma 4.2). This implies that with high probability there exists a vertex in the graph which is not satisfied by any vector in $\text{span}(E)$, and hence with such probability, $E$ is not an index code for the graph.

The following lemma bounds from above the probability that the graph $\vec{G}(n,p)$ has an index code for which many vertices are satisfied by vectors of high Hamming weight.

**Lemma 4.1.** *For every field $\mathbb{F}$ and $n, r, s \in \mathbb{N}$, the probability that there exist a linear $\ell$-index code $E \in \mathbb{F}^{n \times \ell}$ for $\vec{G}(n,p)$ over $\mathbb{F}$ and $s$ vertices, each of which is satisfied by a vector in $\text{span}(E) \setminus B_n(r)$, is at most*

$$\binom{n}{s} \cdot |\mathbb{F}|^{n\ell} \cdot \left(|\mathbb{F}|^{\ell} \cdot p^r\right)^s.$$

**Proof:** Fix a linear $\ell$-index code $E$ for $\vec{G}(n,p)$ over $\mathbb{F}$ and a set $S \subseteq [n]$ of $s$ vertices. The probability that a vertex $i$ is satisfied by a fixed vector $y \in \text{span}(E) \setminus B_n(r)$ is at most $p^r$. To see this, notice that every vertex (except $i$) which corresponds to a nonzero entry of $y$ must be a neighbor of $i$, and this happens independently with probability $p$. Taking the union bound over all the vectors in $\text{span}(E) \setminus B_n(r)$, we get that the probability that a vertex is satisfied by a vector in $\text{span}(E) \setminus B_n(r)$ is at most $|\mathbb{F}|^{\ell} \cdot p^r$. Hence, by the independence of the edges in $\vec{G}(n,p)$, the probability that every vertex in $S$ is satisfied by a vector in $\text{span}(E) \setminus B_n(r)$ is at most $\left(|\mathbb{F}|^{\ell} \cdot p^r\right)^s$. Now, apply the union bound over all the matrices $E$ and sets $S$ to get the desired bound. ∎

Now we turn to deal with vertices which are satisfied by vectors of low Hamming weight and to bound from above their number.

**Lemma 4.2.** *For every field $\mathbb{F}$, a graph $G$, and a linear $\ell$-index code for $G$ over $\mathbb{F}$, at most $\frac{n}{2}$ vertices in $G$ are satisfied by vectors of Hamming weight at most $\frac{n}{2\ell}$.*

**Proof:** Let $E \in \mathbb{F}^{n \times \ell}$ be a generator matrix of a linear $\ell$-index code for $G$ over $\mathbb{F}$. By Claim 2.3, the number of indices of coordinates that are nonzero in at least one vector in $\text{span}(E) \cap B_n(\frac{n}{2\ell})$ is at most $\frac{n}{2}$. Recall that a vector which satisfies a vertex $i$ must have the $i$th entry nonzero. Hence, the number of vertices that can be satisfied by vectors in $\text{span}(E)$ of Hamming weight at most $\frac{n}{2\ell}$ is at most $\frac{n}{2}$. ∎

The $\Omega(\sqrt{n})$ lower bound follows from combining Lemmas 4.1 and 4.2.

**Theorem 4.3.** *For every field $\mathbb{F}$ and $p \in (0,1)$, almost surely*

$$\text{minrk}_{\mathbb{F}}(G(n,p)) = \Omega\left(\sqrt{n} \cdot \sqrt{\frac{\log \frac{1}{p}}{\log |\mathbb{F}|}}\right).$$

**Proof:** Take $\ell \leq \sqrt{n} \cdot \sqrt{\frac{\log \frac{1}{p}}{8 \log |\mathbb{F}|}}$. By Lemma 3.1, it suffices to prove that $\text{minrk}_\mathbb{F}(\vec{G}(n,p)) \geq \ell$ almost surely. Let $G$ be a graph distributed according to $\vec{G}(n,p)$. Let $A$ denote the event that there exist a linear $\ell$-index code $E$ for $G$ over $\mathbb{F}$ and $\frac{n}{2}$ vertices in $G$, each of which is satisfied by a vector in $\text{span}(E) \setminus B_n(\frac{n}{2\ell})$. By Lemma 4.1,

$$\Pr[A] \leq \binom{n}{\frac{n}{2}} \cdot |\mathbb{F}|^{n\ell} \cdot \left(|\mathbb{F}|^\ell \cdot p^{\frac{n}{2\ell}}\right)^{\frac{n}{2}} \leq 2^n \cdot |\mathbb{F}|^{\frac{3}{2}n\ell} \cdot p^{\frac{n^2}{4\ell}} = |\mathbb{F}|^{-\Omega(n\ell)}.$$

On the other hand, by Lemma 4.2, there is no linear $\ell$-index code $E$ for $G$ over $\mathbb{F}$ and $\frac{n}{2}$ vertices in $G$, each of which is satisfied by a vector in $\text{span}(E) \cap B_n(\frac{n}{2\ell})$. This implies that almost surely there is no linear $\ell$-index code for $G$ over $\mathbb{F}$. ∎

## 5 Locally Decodable Index Codes

In this section we study locally decodable index codes defined as follows.

**Definition 5.1.** *A $(q, \ell)$-locally decodable index code is an $\ell$-index code in which the query complexity of the decoding is at most $q$. This means that for every $i$ the decoding function $D_i$ of the ith receiver queries at most $q$ characters from the encoded message.*

**Remark 5.2.** *For every graph $G$, the minimum $\ell$ for which there is a $(1, \ell)$-locally decodable index code for $G$ over $\mathbb{F}$ is the clique cover number $\chi(\overline{G})$ of $G$.*

The following theorem shows a lower bound on the length of a linear locally decodable index code for $G(n,p)$ over $\mathbb{F}$. Although more involved, its proof follows the nature of the proof given for Theorem 4.3.

**Theorem 5.3.** *For every constant size field $\mathbb{F}$ and a constant $p \in (0,1)$, there exist constants $c_1, c_2 > 0$ such that if $\ell \leq c_1 \cdot \frac{n^{2/3}}{(\log n)^{1/3}}$ and $q \leq c_2 \cdot \frac{n}{\ell \cdot \log \ell}$ then almost surely there is no linear $(q, \ell)$-locally decodable index code for $G(n,p)$ over $\mathbb{F}$.*

**Proof:** Let $\ell$ and $q$ be as in the theorem and define $r = \frac{10\ell \cdot \log |\mathbb{F}|}{\log \frac{1}{p}}$. By Lemma 3.1, it suffices to show that almost surely there is no linear $(q, \ell)$-locally decodable index code for $\vec{G}(n,p)$ over $\mathbb{F}$. For a graph $G$ distributed according to $\vec{G}(n,p)$ consider the following two events:

- $A_1$: there exist a linear $\ell$-index code $E$ for $G$ and $\frac{n}{4}$ vertices in $G$, each of which is satisfied by a vector in $\text{span}(E) \setminus B_n(r)$.

- $A_2$: there exist a subspace $W \subseteq \mathbb{F}^n$ and a set $S$ of $\frac{n}{4}$ vertices in $G$ such that

  1. $W$ is spanned by $\ell$ vectors of Hamming weight in $(\frac{n}{2\ell}, r]$, and
  2. there exists a set $U \subseteq W$ of $\ell$ vectors such that every vertex in $S$ is satisfied by a vector which is a linear combination of at most $q$ vectors in $U$ and has Hamming weight greater than $\frac{n}{2\ell}$.

The following lemma reduces the lower bound in the theorem to analyzing the probabilities of $A_1$ and $A_2$.

**Lemma 5.4.** *If there is a linear $(q, \ell)$-locally decodable index code $E$ for $G$ over $\mathbb{F}$ then at least one of $A_1$ and $A_2$ occurs.*

**Proof:** Let $e_1, \ldots, e_\ell \in \mathbb{F}^n$ be the vectors for which every $x \in \mathbb{F}^n$ is mapped by $E$ to $E(x) = (\langle e_1, x \rangle, \langle e_2, x \rangle, \ldots, \langle e_\ell, x \rangle)$. By Lemma 4.2, there are at most $\frac{n}{2}$ vertices in $G$ that are satisfied by vectors in $\text{span}(E)$ of Hamming weight at most $\frac{n}{2\ell}$. Assume that $A_1$ does not occur. This implies that there exists a set $S$ of $\frac{n}{4}$ vertices in $G$ which are satisfied by vectors in $\text{span}(E)$ with Hamming weight in $(\frac{n}{2\ell}, r]$ (and are *not* satisfied by any other vectors in $\text{span}(E)$). Let $y'_1, \ldots, y'_{\frac{n}{4}}$ be vectors that satisfy the vertices in $S$, and let $W$ be their linear span. Notice that $W$ is spanned by $\ell$ of these vectors, thus Item 1 of event $A_2$ holds. Since $W \subseteq \text{span}(E)$, the subspace $\text{span}(E)$ equals the direct sum $W \oplus V$ for some subspace $V \subseteq \mathbb{F}^n$. Thus, for every $j \in [\ell]$, the vector $e_j$ can be uniquely written as $e_j = w_j + v_j$ for some $w_j \in W$ and $v_j \in V$.

We claim that the set $U = \{w_1, \ldots, w_\ell\}$ satisfies the requirement in Item 2 of event $A_2$. Indeed, for every vertex $z \in S$ there exist a set $J \subseteq [\ell]$ of size at most $q$ and coefficients $a_j \in \mathbb{F}$ for $j \in J$ such that $z$ is satisfied by the vector $w_z = \sum_{j \in J} a_j \cdot e_j \in W$ which has Hamming weight in $(\frac{n}{2\ell}, r]$. Notice that $w_z = \sum_{j \in J} a_j \cdot w_j + \sum_{j \in J} a_j \cdot v_j$, where $\sum_{j \in J} a_j \cdot w_j \in W$ and $\sum_{j \in J} a_j \cdot v_j \in V$. Since a vector in $\text{span}(E)$ can be uniquely written as a sum of a vector in $W$ and a vector in $V$, we must have $w_z = \sum_{j \in J} a_j \cdot w_j$. This yields that the vertex $z$ is satisfied by the vector $w_z$ which is a linear combination of at most $q$ vectors in $U$ and has Hamming weight greater than $\frac{n}{2\ell}$. We conclude that $A_2$ occurs, and we are done. ∎

We turn to prove that each of $A_1$ and $A_2$ occurs with probability exponentially small in $n$, and this implies, by the union bound, that there is no linear $(q, \ell)$-index code for $G$ almost surely.

By Lemma 4.1 and the definition of $r$,
$$\Pr[A_1] \leq \binom{n}{\frac{n}{4}} \cdot |\mathbb{F}|^{n\ell} \cdot \left(|\mathbb{F}|^\ell \cdot p^r\right)^{\frac{n}{4}} \leq 2^n \cdot |\mathbb{F}|^{n\ell} \cdot \left(|\mathbb{F}|^\ell \cdot |\mathbb{F}|^{-10\ell}\right)^{\frac{n}{4}} = |\mathbb{F}|^{-\Omega(n\ell)}.$$

To bound from above the probability of $A_2$, we use the union bound over all the subspaces $W$, sets $S$ and sets $U$. The number of subspaces $W$ is at most the number of spanning sets, which is bounded by $(\binom{n}{r} \cdot |\mathbb{F}|^r)^\ell$, by Item 1 of event $A_2$. The number of sets $S$ is $\binom{n}{\frac{n}{4}}$, and the number of sets $U \subseteq W$ of size $\ell$ is at most $|\mathbb{F}|^{\ell^2}$. The probability that a vertex in $S$ is satisfied by a vector of Hamming weight greater than $\frac{n}{2\ell}$ is at most $p^{\frac{n}{2\ell}}$, and we take the union bound over all the vectors that can be written as a linear combination of at most $q$ vectors in $U$, whose number is at most $\binom{\ell}{q} \cdot |\mathbb{F}|^q$. Recall that $r = \Theta(\ell)$ and observe that

$$\begin{aligned} \Pr[A_2] &\leq \left(\binom{n}{r} \cdot |\mathbb{F}|^r\right)^\ell \cdot \binom{n}{\frac{n}{4}} \cdot |\mathbb{F}|^{\ell^2} \cdot \left(\binom{\ell}{q} \cdot |\mathbb{F}|^q \cdot p^{\frac{n}{2\ell}}\right)^{\frac{n}{4}} \\ &\leq 2^{O(\ell^2 \log n + n + \ell^2 + nq \log \ell) - \Omega(\frac{n^2}{\ell})} = 2^{-\Omega(\frac{n^2}{\ell})}, \end{aligned}$$

where the last equality follows from our assumptions on $\ell$ and $q$ for an appropriate choice for $c_1$ and $c_2$. ∎

## 6 Low Density Generator Matrix Index Codes

In this section we study low density generator matrix index codes (or, in short, low density index codes). As will be presented in detail shortly, to obtain our lower bounds (in Section 6.2) we use proof techniques that differ significantly from those previously presented. A formal definition of low density index codes follows.

**Definition 6.1.** *A $(q, \ell)$-low density index code is a linear $\ell$-index code in which every character of the sent word affects at most $q$ characters in the encoded message. Equivalently, it is a linear $\ell$-index code whose generator matrix has at most $q$ nonzero entries in a row.*

**Remark 6.2.** *For every graph $G$, the minimum $\ell$ for which there is a $(1, \ell)$-low density index code for $G$ over $\mathbb{F}$ is the clique cover number $\chi(\overline{G})$ of $G$.*

## 6.1 The Reduction to $q = \omega(1)$

The following theorem shows that in order to prove an $\omega(\sqrt{n})$ lower bound on the minimum length of a linear index code for $G(n, p)$ over a field $\mathbb{F}$, it is enough to prove such a lower bound on the length of a *low density* index code for $G(n, p)$ over $\mathbb{F}$ for *some* $q = \omega(1)$. For simplicity, we state and prove the result for the directed graph $\vec{G}(n, p)$, but using Lemma 3.1 one can obtain a similar result for the undirected random graph $G(n, p)$.

**Theorem 6.3.** *For every field $\mathbb{F}$ and $p \in (0, 1)$, if the probability that $\vec{G}(n, p)$ has a $(q, \ell)$-low density index code over $\mathbb{F}$ is $2^{-\omega(n)}$ for some $q = \omega(1)$ and $\ell = \omega(\sqrt{n})$, then the minimum length of a linear index code for $\vec{G}(n, p)$ over $\mathbb{F}$ is almost surely $\omega(\sqrt{n})$.*

**Proof:** Fix $\mathbb{F}$ and $p \in (0, 1)$. We start by rephrasing the assumption in the theorem statement to allow a more structured proof. Namely, it follows from the theorem's assumption that there exists a non-decreasing function $g : \mathbb{N} \to (0, \infty)$ satisfying $g(n) = \omega(1)$ such that the probability that $\vec{G}(n, p)$ has a $(g(n)^2, \sqrt{n} \cdot g(n))$-low density index code over $\mathbb{F}$ is at most $2^{-8n}$. We will prove that almost surely there is no linear $\ell$-index code for $\vec{G}(n, p)$ over $\mathbb{F}$ for $\ell = \sqrt{n} \cdot f(n)$, where $f : \mathbb{N} \to (0, \infty)$ is the function defined by $f(n) = \min(\frac{1}{2}, \sqrt{\frac{\log \frac{1}{p}}{16 \log |\mathbb{F}|}}) \cdot g(\lceil \frac{n}{4} \rceil)$. Notice that $\ell = \omega(\sqrt{n})$ and hence the theorem will follow.

Let $G$ be a graph distributed according to $\vec{G}(n, p)$, and consider the following two events:

- $A_1$: there exists a set of $\frac{n}{4}$ vertices in $G$ whose induced subgraph has a $(\frac{16 f(n)^2 \cdot \log |\mathbb{F}|}{\log \frac{1}{p}}, \ell)$-low density index code.

- $A_2$: there exist a linear $\ell$-index code $E$ for $G$ and $\frac{n}{2}$ vertices in $G$, each of which is satisfied by a vector in $\text{span}(E) \setminus B_n(\frac{4\ell \cdot \log |\mathbb{F}|}{\log \frac{1}{p}})$.

First, we claim that every graph $G$ that has a linear $\ell$-index code must satisfy at least one of the events $A_1$ and $A_2$. To see why, consider a graph $G$ that has a linear $\ell$-index code $E \in \mathbb{F}^{n \times \ell}$ and does not satisfy $A_2$. Observe that $G$ has a set $S$ of $\frac{n}{2}$ vertices that are satisfied by vectors in $\text{span}(E)$ of Hamming weight at most $\frac{4\ell \cdot \log |\mathbb{F}|}{\log \frac{1}{p}}$. Take a maximal linearly independent subset of the $\frac{n}{2}$ vectors which satisfy the vertices in $S$ and restrict them to the coordinates that correspond to vertices in $S$. The matrix with these restricted vectors as columns has at most $\ell$ columns and consists of at most $\ell \cdot \frac{4\ell \cdot \log |\mathbb{F}|}{\log \frac{1}{p}} = \frac{4n \cdot f(n)^2 \cdot \log |\mathbb{F}|}{\log \frac{1}{p}}$ nonzero entries. This implies that this matrix has at least $\frac{n}{4}$ rows each of which has at most $\frac{16 f(n)^2 \cdot \log |\mathbb{F}|}{\log \frac{1}{p}}$ nonzero entries. Restricting the matrix to these $\frac{n}{4}$ rows, we get that $A_1$ holds.

Now, we turn to bound from above the probabilities of the events $A_1$ and $A_2$. Clearly, every induced subgraph of $\vec{G}(n, p)$ on $\frac{n}{4}$ vertices is distributed according to $\vec{G}(\frac{n}{4}, p)$. Therefore, the probability that it has a $(g(\lceil \frac{n}{4} \rceil)^2, \frac{\sqrt{n}}{2} \cdot g(\lceil \frac{n}{4} \rceil))$-low density index code over $\mathbb{F}$ is at most $2^{-8 \cdot \frac{n}{4}} = 2^{-2n}$. Using the definition of $f$ and the union bound taken over the subsets of $[n]$ of size $\frac{n}{4}$, we obtain

$$\Pr[A_1] \leq \binom{n}{\frac{n}{4}} \cdot 2^{-2n} \leq 2^{-n}.$$

By Lemma 4.1,
$$\Pr[A_2] \leq \binom{n}{\frac{n}{2}} \cdot |\mathbb{F}|^{n\ell} \cdot \left(|\mathbb{F}|^{\ell} \cdot p^{\frac{4\ell \cdot \log |\mathbb{F}|}{\log \frac{1}{p}}}\right)^{\frac{n}{2}} = |\mathbb{F}|^{-\Omega(n\ell)}.$$

By the union bound the probability that at least one of the events occurs is smaller than $2^{-\Omega(n)}$, thus, with such probability, $\vec{G}(n,p)$ has a linear $\ell$-index code. ∎

## 6.2 The Lower Bounds for $q \in \{2, 3\}$

The following theorem says that every index code for $G(n,p)$ whose generator matrix has at most 3 nonzero entries in a row has length $\omega(\sqrt{n})$.

**Theorem 6.4.** *For every field $\mathbb{F}$ and a sufficiently small $\varepsilon > 0$ there exists a $p' = p'(|\mathbb{F}|, \varepsilon) > 0$ such that for any $p \in (0, p')$ the following holds almost surely.*

1. *If there is a $(2, \ell)$-low density index code for $G(n,p)$ over $\mathbb{F}$ then $\ell \geq n^{1-\varepsilon}$.*
2. *If there is a $(3, \ell)$-low density index code for $G(n,p)$ over $\mathbb{F}$ then $\ell \geq n^{\frac{2}{3}-\varepsilon}$.*

The approach in the proof of Theorem 6.4 is different from the one taken in the previous proofs. As before, fix a linear index code for $\vec{G}(n,p)$ and denote its generator matrix by $E \in \mathbb{F}^{n \times \ell}$. Assume that every row of $E$ consists of at most $q$ nonzero entries for $q \in \{2,3\}$. Let $i$ be a vertex and let $A$ denote the set of rows in $E$ that correspond to non-neighbors of $i$. We are asking if there exists a vector $v \in \text{span}(E)$ which satisfies $i$ (i.e., $v_i \neq 0$ and $v_j = 0$ for all $j \neq i$ to which $i$ is not connected). One can show that there exists such a vector if any only if the $i$th row of $E$ cannot be written as a linear combination of a subset of the rows in $A$. The reason is that such a subset of rows enforces any vector in $\text{span}(E)$ all of whose entries corresponding to rows in $A$ are zero, to have zero in the $i$th entry as well, and in particular not to satisfy $i$.

In our proof we show that every matrix $E \in \mathbb{F}^{n \times \ell}$ has many small sets $F$ of rows which are minimally linearly dependent (where minimality is with respect to containment). As we will show later on, this can be achieved using the assumption that $E$ has low density (at most $q$ nonzero entries in a row). Notice that if the $i$th row of $E$ belongs to such $F$, then the $i$th row of $E$ can be written as a linear combination of $|F| - 1$ rows of $E$. If all the vertices that correspond to these rows are non-neighbors of $i$ then $i$ has no satisfying vector in $\text{span}(E)$. Therefore, the probability that $i$ has a satisfying vector is at most $1 - (1-p)^{|F|-1}$.

Our construction of minimally linearly dependent row sets of $E$ is based on a result of Naor and Verstraëte [20]. They studied the maximum size of a set of vectors in $\mathbb{F}^N$ with Hamming weight at most $q$ in which every subset of size $k$ is linearly independent over $\mathbb{F}$. For $\mathbb{F}_2$, this is the minimum number of edges of size at most $q$ in a hypergraph on $N$ vertices which does not contain an even cover[4] of size at most $k$. We now add the notion of a dependence set and use it to state the result of [20]. We note that we use their result only for $q \in \{2, 3\}$, as for larger $q$ our approach does not improve upon the $\Omega(\sqrt{n})$ bound given in Theorem 4.3.

**Definition 6.5.** *A subset of $\mathbb{F}^N$ is a k-dependence set if it is a linearly dependent set over $\mathbb{F}$ whose size is at most k.*

**Theorem 6.6** ([20]). *For every field $\mathbb{F}$, $q \in \mathbb{N}$ and $k \geq 8$, there exists a constant $c = c(|\mathbb{F}|, q, k) > 0$ for which the following holds for every $N \in \mathbb{N}$. Every subset[5] of $\mathbb{F}^N$ of at least $c \cdot N^{\frac{q}{2} + \frac{\lceil q/3 \rceil}{2\lceil k/8 \rceil}}$ vectors with Hamming weight at most q, contains a k-dependence set.*

---

[4]An *even cover* is a non-empty set of edges such that every vertex belongs to an even number of them.
[5]Throughout this section we allow multiplicities in the vector sets.

Equipped with Theorem 6.6 we are ready to state and prove a lemma on the existence of many small dependence sets in a given set of vectors. The basic idea in constructing these sets is to apply Theorem 6.6 iteratively. However, it turns out (as will be explained later) that in order to avoid dependencies in our probability analysis we need every two dependence sets of the construction to share at most one element. In what follows we add the notion of 1-intersecting family of sets and then state and prove our lemma.

**Definition 6.7.** *A family $\mathcal{F}$ of sets is* 1-intersecting *if every distinct $A, B \in \mathcal{F}$ share at most one common element.*

**Lemma 6.8.** *For every field $\mathbb{F}$, $q \in \mathbb{N}$, a sufficiently small $\varepsilon > 0$ and a sufficiently large $N \in \mathbb{N}$ the following holds. For every set $A \subseteq \mathbb{F}^N$ of at least $N^{\frac{q}{2}+2\varepsilon}$ vectors with Hamming weight at most $q$, there exists a 1-intersecting family $\mathcal{F} \subseteq P(A)$ of k-dependence sets for some $k \leq \frac{5q}{\varepsilon}$ that satisfies*

$$|\mathcal{F}| \geq \Omega\left(N^{\frac{q}{2}+2\varepsilon} \log N \cdot \frac{\varepsilon}{k \log k}\right).$$

**Proof:** Let $A$ be a set of vectors in $\mathbb{F}^N$ with Hamming weight at most $q$ and assume that $|A| \geq N^{\frac{q}{2}+2\varepsilon}$. Let $k$ be the smallest integer so that $\varepsilon \geq \frac{\lceil q/3 \rceil}{2\lceil k/8 \rceil}$, and notice that $k \leq \frac{5q}{\varepsilon}$ for any small enough $\varepsilon$.

We construct a family $\mathcal{F} \subseteq P(A)$ of $k$-dependence sets as follows. Start with $\mathcal{F} = \phi$ and $A' = A$. As long as $|A'| \geq N^{\frac{q}{2}+\frac{3}{2}\varepsilon}$, add to $\mathcal{F}$ a $k$-dependence set $F \subseteq A'$, whose existence is guaranteed by Theorem 6.6 using our choice of $k$, and continue with $A' \setminus F$. Notice that in this way we collect at least $\frac{|A|}{2k}$ $k$-dependence sets. Now, partition $A$ into $k$ sets $A_1, \ldots, A_k$ of size $\frac{|A|}{k}$ each, so that no $F$ which was added to $\mathcal{F}$ in the previous step shares more than one element in common with some $A_i$. To achieve this, partition the elements of every such $F$ into the $k$ sets, at most one element in each $A_i$, and then partition the remaining elements in a way that all the $A_i$'s have (roughly) the same size. We continue recursively with the $A_i$'s and add the new $k$-dependence sets to the same $\mathcal{F}$.

In the second iteration of the recursion we get at least $\frac{|A_i|}{2k} \geq \frac{|A|}{2k^2}$ new $k$-dependence sets from every $A_i$, so their total number is at least $\frac{|A|}{2k}$. In the $i$th iteration of the recursion there are $k^{i-1}$ sets of size $\frac{|A|}{k^{i-1}}$ each. Each of them contributes at least $\frac{|A|}{2k^i}$ $k$-dependence sets to $\mathcal{F}$ and their total contribution is at least $\frac{|A|}{2k}$. Notice that the recursion does not terminate as long as the sets are of size at least $N^{\frac{q}{2}+\frac{3}{2}\varepsilon}$, and hence the recursion depth is at least $\log_k N^{\frac{\varepsilon}{2}}$. In each iteration we add to $\mathcal{F}$ at least $\frac{|A|}{2k}$ $k$-dependence sets, so the final $\mathcal{F}$ satisfies

$$|\mathcal{F}| = \Omega\left(\frac{|A|}{2k} \cdot \log_k N^{\frac{\varepsilon}{2}}\right) = \Omega\left(N^{\frac{q}{2}+2\varepsilon} \log N \cdot \frac{\varepsilon}{k \log k}\right).$$

Finally, observe that in every level of the recursion we get disjoint $k$-dependence sets. Also, notice that the recursion is always applied to sets with intersection size at most 1 with every $k$-dependence set that was previously added to $\mathcal{F}$. This implies that $\mathcal{F}$ is 1-intersecting. ∎

Now we turn to prove Theorem 6.4.

**Proof of Theorem 6.4:** Let $q \in \{2, 3\}$, fix a $(q, \ell)$-low density index code for $G(n, p)$ over $\mathbb{F}$ with $\ell = n^{\frac{2}{q}-\varepsilon}$, and denote its generator matrix by $E \in \mathbb{F}^{n \times \ell}$. The number of nonzero entries in a row of $E$ is at most $q$. Let $A$ be the set of rows of $E$ (possibly with multiplicities). This is a set of vectors in $\mathbb{F}^\ell$ with Hamming weight at most $q$. Notice that $|A| = n = \ell^{\frac{1}{\frac{2}{q}-\varepsilon}} = \ell^{\frac{q}{2}+2\varepsilon'}$ for some $\varepsilon' > \frac{\varepsilon}{2}$. By Lemma 6.8, there is a 1-intersecting family $\mathcal{F} \subseteq P(A)$ of $k$-dependence sets for some $k \leq \frac{5q}{\varepsilon'} \leq \frac{15}{\varepsilon'} \leq \frac{30}{\varepsilon}$ such that $|\mathcal{F}| \geq \Omega(n \log \ell \cdot \frac{\varepsilon^2}{\log \frac{1}{\varepsilon}})$. Assume, without loss of generality, that the sets in $\mathcal{F}$ are minimal (i.e., do not contain any proper linearly dependent subset). For simplicity, let us think of every set $F \in \mathcal{F}$ as a subset of $[n]$ that consists of the indices of the rows in $F$.

For every $k$-dependence set $F \in \mathcal{F}$ and $i \in F$, the vertex $i$ must be connected to a least one of the other vertices in $F$. Otherwise, a satisfying vector of $i$ has zeros in all the entries with indices in $F \setminus \{i\}$. Such a vector must have a zero in the $i$th entry as well, since the row that corresponds to vertex $i$ can be written as a linear combination of the other rows in $F$. This yields that the vector does not satisfy $i$. Therefore, with probability $(1-p)^{|F|-1} \geq (1-p)^{k-1}$ the vertex $i$ is not satisfied by any vector in $\text{span}(E)$.

Now, we apply this argument to every set $F \in \mathcal{F}$ and an arbitrarily chosen vertex $i \in F$ and we bound from above the probability that $i$ is satisfied by a vector in $\text{span}(E)$. Observe that these events are independent, since if $i$ is the vertex that was chosen from the sets $F_1$ and $F_2$ then the sets $F_1 \setminus \{i\}$ and $F_2 \setminus \{i\}$ are disjoint because $\mathcal{F}$ is 1-intersecting. So the probability that every vertex in the graph is satisfied by a vector in $\text{span}(E)$ is at most $(1 - (1-p)^{k-1})^{|\mathcal{F}|}$. Taking the union bound over at most $\left(\binom{\ell}{q} \cdot |\mathbb{F}|^q\right)^n$ generator matrices with at most $q$ nonzero entries in a row, we obtain that the probability that there exists a $(q, \ell)$-low density generator matrix index code for $G(n, p)$ over $\mathbb{F}$ is at most

$$\left(\binom{\ell}{q} \cdot |\mathbb{F}|^q\right)^n \cdot (1 - (1-p)^{k-1})^{|\mathcal{F}|} \leq 2^{qn(\log \ell + \log |\mathbb{F}|)} \cdot (1 - (1-p)^{\frac{30}{\varepsilon}})^{\Omega\left(n \log \ell \cdot \frac{\varepsilon^2}{\log \frac{1}{\varepsilon}}\right)}.$$

To complete the proof, notice that for any small enough $p$ (depending only on $|\mathbb{F}|$ and $\varepsilon$) the above tends exponentially to zero as $n$ tends to infinity. ∎

## 7 Concluding Remarks and Open Questions

In this paper we initiated the study of index coding for the random graph $G(n, p)$ over a field $\mathbb{F}$ and introduced two new models of index coding – locally decodable index coding and low density index coding. We proved several lower bounds on the length of linear index codes for $G(n, p)$ (Theorems 4.3, 5.3, 6.4) and showed that in order to improve the $\Omega(\sqrt{n})$ lower bound it suffices to improve it for low density index codes (Theorem 6.3).

The main task left for further work is to obtain tighter bounds on the minimum length of index codes for the random graph $G(n, p)$ over a field $\mathbb{F}$. More specifically, it is an open question if there exists an index code for $G(n, p)$ (linear or not) shorter than the one achieved by the clique cover. It is interesting if our lower bounds can be extended to general (non-linear) index codes. It would be nice to understand better how the limit on the number of queries affects the length of locally decodable index codes for $G(n, p)$. We hope that the new notion of low density index codes and Theorem 6.3 will be found useful in understanding the minrank of $G(n, p)$ over $\mathbb{F}$.

Another challenging research direction is to study the *vector* capacity of the random graph $G(n, p)$ (see [18, 3, 7]). Here, the sender wishes to broadcast a word $x$ of $n$ blocks, each of $t$ bits, to $n$ receivers. The $i$th receiver is interested in the $i$th block and has side information consisting of a subset of the other blocks according to $G(n, p)$. Denoting by $\beta_t$ the minimum number of bits that has to be transmitted, we are interested in $\lim_{t \to \infty} \frac{\beta_t}{t}$. This limit represents the average communication cost per bit in each block (for long blocks), and it will be very interesting to compare it to $\beta_1$ of a typical random graph.